\documentclass[
 reprint,
 superscriptaddress,
 amsmath,amssymb,
 aps,
 prl,
 secnumarabic,
 longbibliography,
]{revtex4-2}
\usepackage{comment}
\usepackage{graphicx}
\usepackage{dcolumn}
\usepackage{bm}
\usepackage{xcolor}
\usepackage{xr}
\usepackage[T1]{fontenc}

\begin{document}

\preprint{APS/123-QED}

\title{Localization of an Individual Charge Fluctuator in a Si/SiGe Spin Qubit Device}

\author{Leah Tom}
 \affiliation{Department of Electrical and Computer Engineering, University of California, Los Angeles}
\author{John Nichol}
\affiliation{Department of Physics and Astronomy, University of Rochester}
\affiliation{University of Rochester Center for Coherence and Quantum Science}
\author{Chris Anderson}
\affiliation{Department of Mathematics, University of California, Los Angeles}
\author{Mark F. Gyure}
\affiliation{Department of Electrical and Computer Engineering, University of California, Los Angeles}

\date{\today}

\begin{abstract}

Charge noise from two-level fluctuators (TLFs) limits gate fidelities in semiconductor spin qubits, yet locating individual fluctuators within a device remains challenging. Here, we demonstrate a spatial triangulation method that combines the gate dependence of a TLF’s detuning with the chemical potential shifts it induces in a sensing dot. Applying this method to previously reported measurements from a gated Si/SiGe quantum dot device, we use electrostatic simulations to jointly constrain the fluctuator’s three-dimensional position and the magnitude of its electric dipole moment. We localize the fluctuator to a region spanning 31 nm by 1 nm laterally and 4 nm vertically within the bulk of the atomic layer deposited aluminum oxide, identifying the material hosting this individual charge-noise source. The measurements are consistent with a vertically oriented TLF with a magnitude of 3 to 4 eÅ. This approach provides a route to determining which materials and interfaces host active charge fluctuators in semiconductor spin qubit devices.

\end{abstract}

\maketitle

\section{Introduction}

\begin{figure}[t] 
    \centering
    \includegraphics[width=0.40\textwidth]{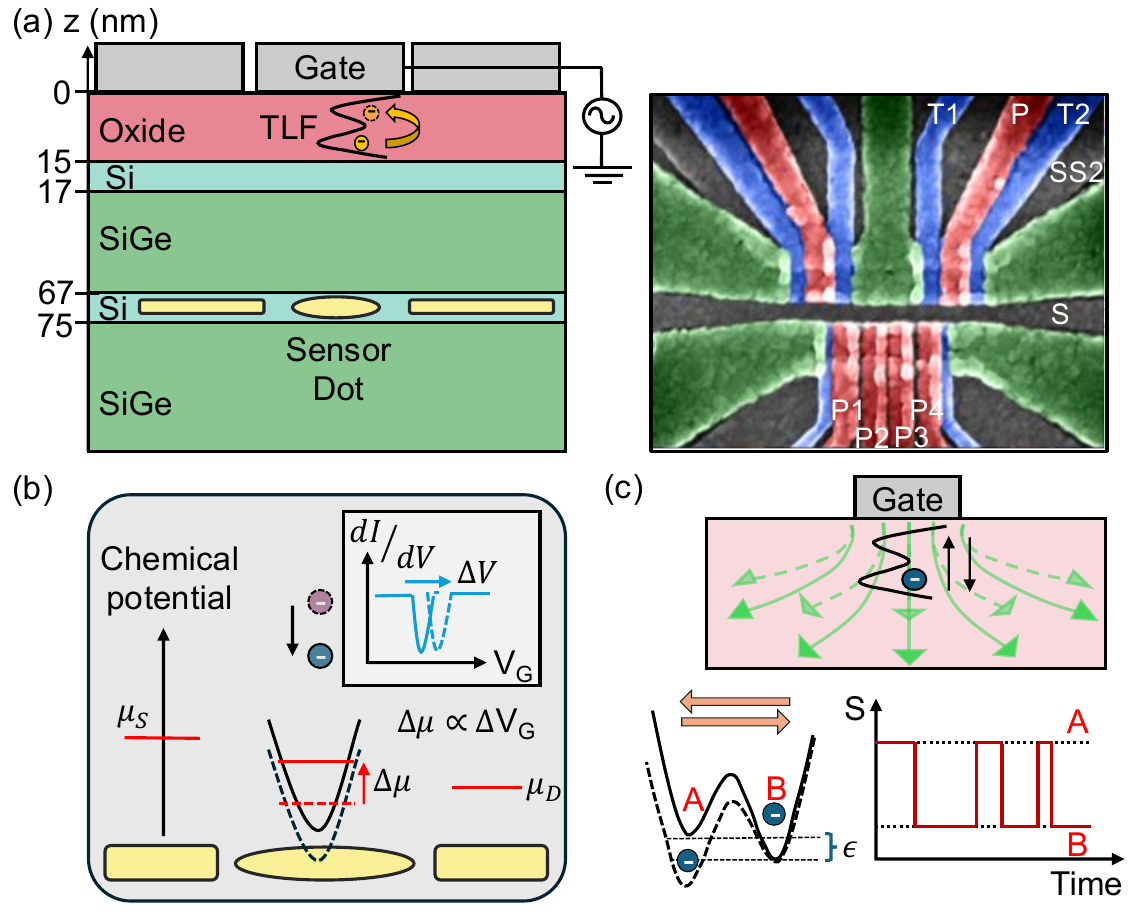}
    \caption[] { (a) Cross-sectional schematic of the simulated Si/SiGe device. The electron density in the silicon quantum well is controlled electrostatically by the metal gates, with the sensing quantum dot located beneath a plunger gate. (b) Measurement of the gate-referred voltage shift of a TLF using RF reflectometry of the sensing dot. The sensing dot is biased on the flank of a Coulomb blockade peak so that the TLF-induced conductance change can be converted to an equivalent plunger gate voltage shift. (c) Measurement of the TLF detuning lever arm. A change in gate voltage modifies the electric field at the TLF, changing the detuning between the TLF's two potential minima. The resulting change in the relative occupation lifetimes is used to extract the gate-TLF detuning lever arm. }
    \label{fig:first_figure}
\end{figure}

Charge noise remains one of the central obstacles to achieving high-fidelity gate operations in semiconductor spin qubits \cite{yoneda_quantum_dot_2018}. Measurements of charge noise spectra in Si/SiGe quantum dots have revealed that low-frequency charge noise can be dominated by a small number of nearby two-level fluctuators (TLFs), with each qubit experiencing a distinct ensemble of fluctuators despite some degree of spatial correlation \cite{connors_low_frequency_2019,rojas_arias_inferring_2025,dumoulin_stuyck_silicon_2024}. To mitigate these effects, several approaches have been developed, including dynamical decoupling sequences, operation at noise-insensitive “sweet spots” \cite{Reed_exchange_symmetricOperation,Russ_Burkard_REX,Shim_Tahan_AEON} and more recently, active stabilization of individual charge fluctuators \cite{ye_stabilizing_2025}. A key step toward more effective noise suppression is the ability to locate or constrain the position of dominant TLFs, which could inform device growth strategies or guide targeted stabilization of problematic defects \cite{ye_stabilizing_2025}. Measurements across devices have linked reduced charge noise to thinner gate dielectrics \cite{connors_low_frequency_2019}, improved semiconductor-dielectric interfaces and improved structural quality achieved by reducing the silicon quantum well thickness \cite{paquelet_wuetz_reducing_2023}. Although candidate locations for defect formation in these devices have been proposed \cite{varley_dangling_2023,Hua2005_threadingdislocation,Lee2003_strainedSi,Simeon2006}, precisely determining the three dimensional locations of an individual TLF remains challenging. Noise correlations provide spatial constraints on individual TLFs, but quantitative location estimates depend on assumptions about the TLF's dipole orientation, magnitude and depth \cite{rojas_arias_inferring_2025}. Related electrostatic triangulation methods have also been used to locate individual donors in silicon \cite{Asaad2020}.

In this work, we demonstrate a spatial triangulation method for locating individual charge fluctuators in three dimensions in semiconductor spin qubit devices. By comparing experimentally measured gate-TLF detuning lever arms and gate-referred voltage shifts \cite{ye_characterization_2024} against high-resolution electrostatic device simulations, we constrain the orientation and dipole-moment magnitude of a TLF and localize it to a region that is 1 nm (x) by 31 nm (y) by 4 nm (z) within the bulk of the atomic layer deposition (ALD) aluminum oxide. The measurements also rule out the charge-trap model considered in Appendix~\ref{app:chargeTrap}. This approach enables quantitative localization of individual charge fluctuators using experimentally accessible measurements, providing a route toward identifying the material origins of charge noise and guiding targeted noise mitigation in semiconductor spin qubit devices.

Improvements in simulation methods, together with multiple high-confidence measurements of the same TLF, enabled the spatial triangulation scheme implemented here. Earlier attempts to locate TLFs or compare simulated TLFs with experimentally measured noise spectra have focused on correlations between noise spectra measured at different quantum dots or on changes in a quantum dot's chemical potential induced by the TLF \cite{rojas_arias_inferring_2025,kepa_simulation_2023}. We demonstrate that just using the chemical potential shift leads to an underconstrained problem and that multiple TLF locations, orientations and dipole moments are consistent with the same chemical potential shift. Here, we use multiple gate-TLF detuning lever arms determined experimentally from time-domain reflectometry measurements, in which the relative lifetimes of the two TLF states change in response to applied gate voltages \cite{ye_characterization_2024}. These additional measurements provide independent constraints on the TLF response, enabling spatial triangulation rather than localization based on a single experimental signature. On the simulation side, locally adaptive meshing and interpolation of the smoothly varying simulated observables enabled an efficient search over TLF configurations while maintaining the numerical accuracy required (Appendix \ref{app:sim_methodology}) . 

\section{Simulation Framework}
\begin{figure}[htbp]
    \centering
    \includegraphics[width=0.48\textwidth]{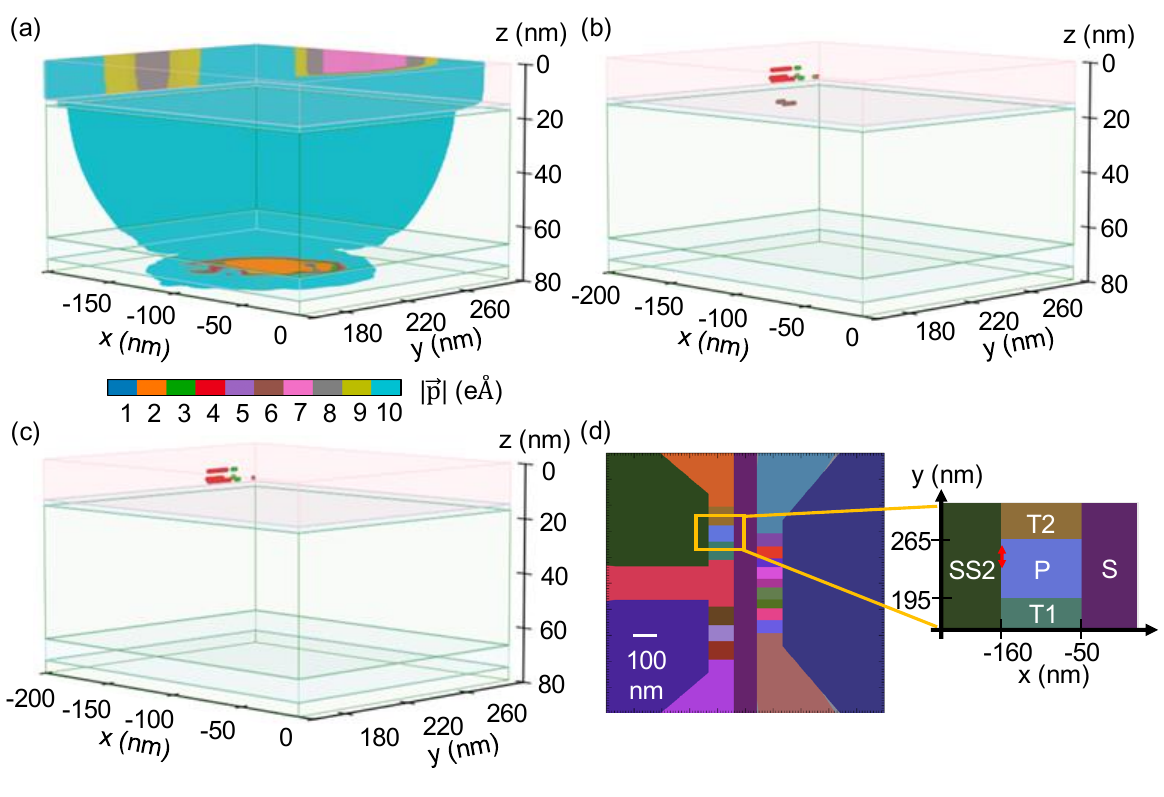}
    \caption[] { (a) The positions and dipole moments consistent with the gate-referred voltage shift of the TLF relative to gate P are plotted. The gate-referred voltage shift by itself cannot be used to constrain the TLF location or dipole moment. (b) The locations and dipole moments consistent with the gate-TLF detuning lever arms with respect to gates P, T1, T2, S and SS2 are plotted. There was only agreement between the 5 datasets for a vertically oriented TLF with electric dipole moments of 3, 4 and 6 e$\mathring{A}$. Based on the detuning lever arm data alone, the TLF is localized to the oxide and top of the SiGe spacer. (c) The final triangulated locations are constrained by both the gate-referred voltage shift data and the detuning lever arm data of all 5 gates. The TLF is localized to a depth of 6 to 10 nm from the surface, within the ALD aluminum oxide and its possible location in the xy plane spans 31 nm in the y-direction and 1 nm in the x-direction. (d) The gate layout at the oxide-gate interface is plotted. The search space in the xy direction is demarcated in yellow by a box. The triangulated location of the TLF is indicated by a red and double-sided arrow in the zoomed in plot of the search space. 
    }
    \label{fig:results_fig_device_layout}
\end{figure}

We simulated the exact device used in recent experimental characterization of a single TLF \cite{ye_characterization_2024} using self-consistent Schrodinger-Poisson simulations. The gate geometry is shown in Fig. \ref{fig:first_figure}. We simulate a Si/SiGe heterostructure stack, with the following parameters. The quantum well is 8 nm wide, the SiGe spacer is 50 nm wide, the silicon cap is 2 nm wide and the ALD aluminum oxide is 15 nm wide. The device was tuned so that there is 1 electron under P3, 0 electrons under P4 and 18 electrons under the charge sensing dot, P. These were the settings chosen for consistency with experiment. The density of electrons under the accumulation gates is determined using a semiclassical method where the electron density is represented as a product of the density of 1D single particle states in the growth direction, z and 2D Thomas-Fermi approximation in the horizontal xy plane. 

The TLF is assumed to correspond to a defect switching between two stable charge configurations and so is modelled as a point charge switching between two closely spaced locations. While the change in charge configurations may not physically correspond to a single charge moving, the calculated multipole moment of the changing charge configuration is approximated to that of a monopole moving between two locations, consistent with the treatment adopted in similar prior work \cite{kepa_simulation_2023,Sarabi2016}. We considered candidate defect locations in proximity to the P gate, to which the TLF is most strongly coupled. Specifically, candidate TLF locations were sampled on a 1 nm grid, spanning x from -200 nm to -10 nm and y from 160 nm to 300 nm. This region is demarcated in yellow in Fig. \ref{fig:results_fig_device_layout}(d). Along the z-direction, candidate TLF locations were evaluated at 1 nm intervals from z = 0 nm to 79 nm, spanning the oxide, silicon cap, SiGe spacer and the SiGe immediately below the well, while excluding the quantum well and the interfaces between layers. There were 3 TLF orientations that were explored along the x, y and z directions. Finally, the magnitude of the monopole displacements varied from 1 $\mathring{A}$ to 10 $\mathring{A}$. This is a proxy for the electric dipole moment of the defect and is a range consistent with what has been measured experimentally \cite{PhysRevApplied_17_034025}.

To determine the TLF location, we determine TLF configurations consistent with the chemical potential shift of the sensing dot induced by the TLF and the gate-TLF detuning lever arms with respect to the five most strongly coupled gates. These six datasets were reported in prior work \cite{ye_characterization_2024} and enable us to localize the TLF to a depth of 6 to 10 nm within the ALD aluminum oxide layer of the device, as shown in Fig. \ref{fig:results_fig_device_layout}. The gate-referred voltage shift is determined experimentally by tuning the sensing dot to be in the Coulomb blockade regime. Specifically, the dot is tuned to be on the flank of a conductance peak as shown in \ref{fig:first_figure}(b). At this position, the conductance is highly sensitive to small potential shifts from the TLF\cite{ye_stabilizing_2025,connors_low_frequency_2019}. The change in chemical potential at the sensing dot changes the conductance of the dot, which is measured using RF reflectometry of the sensing dot. The amplitude of the reflectometry signal is converted using the calibrated plunger-gate lever arm $\alpha_{\mathrm{P}}$ and yields the equivalent gate-voltage change, $
\Delta V_{\mathrm{g,eq}} = \frac{\Delta \mu_{\mathrm{sensor}}}{e\,\alpha_{\mathrm{P}}}.$
This $\Delta V_{\mathrm{g,eq}}$ represents the gate-referred voltage shift—the effective gate-voltage perturbation that would produce the same electrostatic change as the fluctuating TLF~\cite{ye_characterization_2024}. The plunger-gate lever arm, $\alpha_{p}$ was experimentally determined to be 0.053 eV/V. In our simulations, we determine the change in chemical potential at the sensing dot due to the fluctuating TLF and match this to an experimentally determined $\Delta\mu$.

The gate-TLF detuning lever arm was determined by analyzing the relative switching rates of its two states as a function of gate voltage. Time-domain reflectometry traces of the TLF were used to extract the lifetime of the TLF in each state, yielding the transition rates $\Gamma_{A\to B}$ and $\Gamma_{B\to A}$. Their ratio obeys the detailed-balance relation 
$
\frac{\Gamma_{A\to B}}{\Gamma_{B\to A}} = \exp\!\left[-\frac{\Delta E}{k_{\mathrm{B}} T_{\mathrm{eff}}}\right]$,
where $\Delta E$ is the energy difference between the TLF states and $T_{\mathrm{eff}}$ is the effective local temperature. $\Delta E = \sqrt{t^2 + \epsilon^2 }$, where $\epsilon$ is detuning and $t$ is tunnel coupling, for small values of detuning. From this relation, $\epsilon$ was inferred at each applied gate voltage, and a linear fit of $\epsilon$ versus gate voltage offset provided the detuning lever arm, $\lambda = \frac{\partial \epsilon}{\partial V_{\mathrm{g}}}$. This parameter quantifies how sensitively the TLF detuning responds to the coupled gates and can be matched with simulated detuning lever arms. 
\begin{figure}[htbp]
    \centering
    \includegraphics[width=0.45\textwidth]{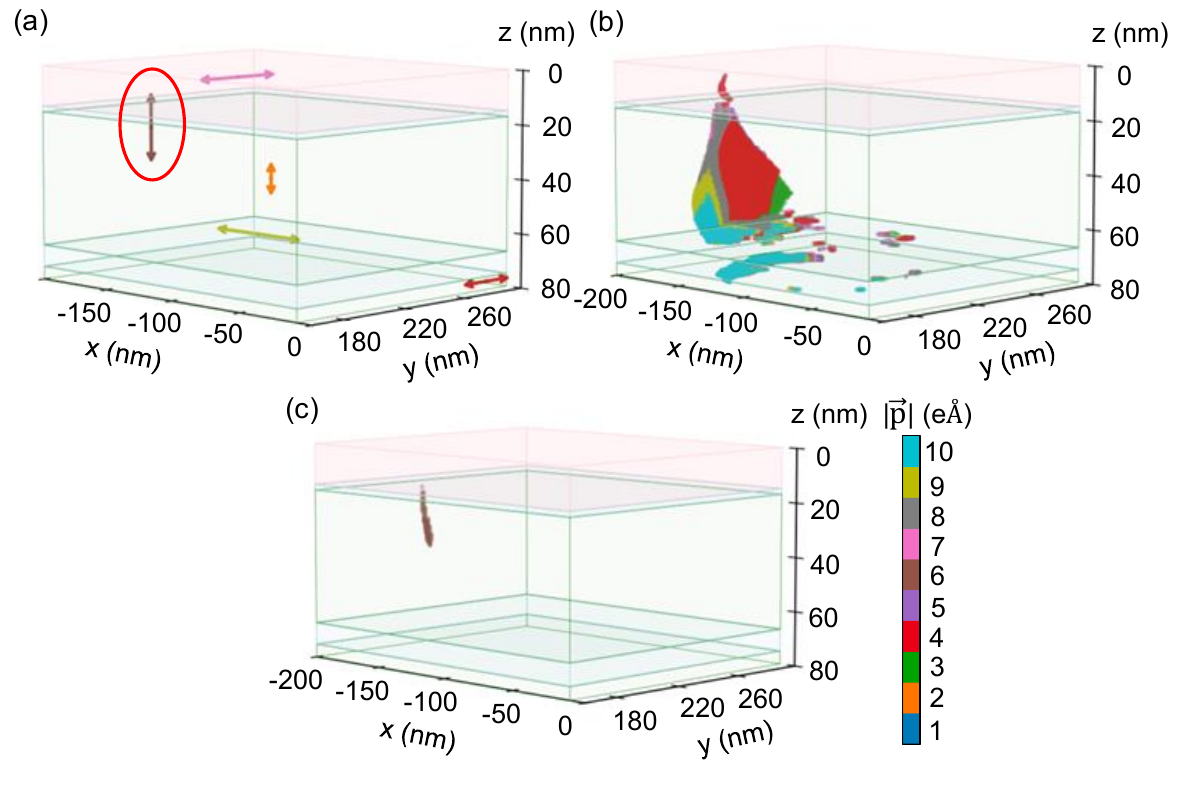}
    \caption[] {Validation of the localization procedure using synthetic fluctuators and its sensitivity to measurement uncertainty. (a) The locations and orientations of a few representative synthetic fluctuators are depicted. These fluctuators sample three orthogonal orientations, different dipole moments and locations in the simulation volume. (b) The results of the triangulation for the vertically oriented fluctuator circled in red are shown. This TLF has an electric dipole moment of 6 e$\mathring{A}$. The triangulation results here correspond to a higher experimental uncertainty in the measured gate-referred voltage shift and detuning lever arms. (c) The triangulation results here for the fluctuator circled in red in (a) correspond to a reduced experimental uncertainty in the gate-referred voltage shift and detuning lever arms in the simulated data.}
    \label{fig:validation_study}
\end{figure}

From simulations, we can determine candidate TLF configurations that give detuning lever arms that are in agreement with experiment. This is done by the following procedure. For a given TLF location, $ \vec{r}$, and electric dipole moment $ \vec{d}$, the simulated gate voltage is reduced by 1 mV for each gate for which we are determining the detuning lever arm. The electron is then placed at the location, $ \vec{r} - \frac{\vec{d}}{2} $ and the environmental potential (see Appendix \ref{app:sim_methodology}) at that location, $\phi_{r-d/2}$ is determined. This is then repeated at a shifted position of $ \vec{r} + \frac{\vec{d}}{2} $ and the environmental potential is recorded as $\phi_{r+d/2}$. The proxy for TLF detuning at this gate voltage is then determined by
\[
 \epsilon_{V_{0}-\delta V}= \phi_{r+d/2} - \phi_{r-d/2} .
\]
Here, $\delta V$= 1 mV. This is repeated for the case where the gate voltage is set to 1 mV above the tuned gate voltage. The simulated gate-TLF detuning lever arm, $\lambda_{sim}$ is determined by,
\[
\lambda_{sim} = \frac{ \epsilon_{V_{0}+\delta V} - \epsilon_{V_{0}-\delta V}}{2\delta V} 
.\]
The simulated gate-TLF detuning lever arm is then compared with experimental measurements of the corresponding gates. The experimental detuning lever arms of the gates we compared against include the gates labelled P, T2, T1, S and SS2, as shown in Fig. \ref{fig:results_fig_device_layout}(d). The detuning lever arms for P, T2, T1, S and SS2 are measured to be $-5.6^{+0.6}_{-0.4}$, $1.33^{+0.27}_{-0.18}$, $0.24^{+0.85}_{-0.60}$, $-0.03^{+0.75}_{-0.43}$ and $0.005^{+0.048}_{-0.052}$ $\mu$eV/mV respectively (95\% confidence interval bounds)\cite{ye_characterization_2024}.

\begin{figure}[htbp]
    \centering
    \includegraphics[width=0.46\textwidth]{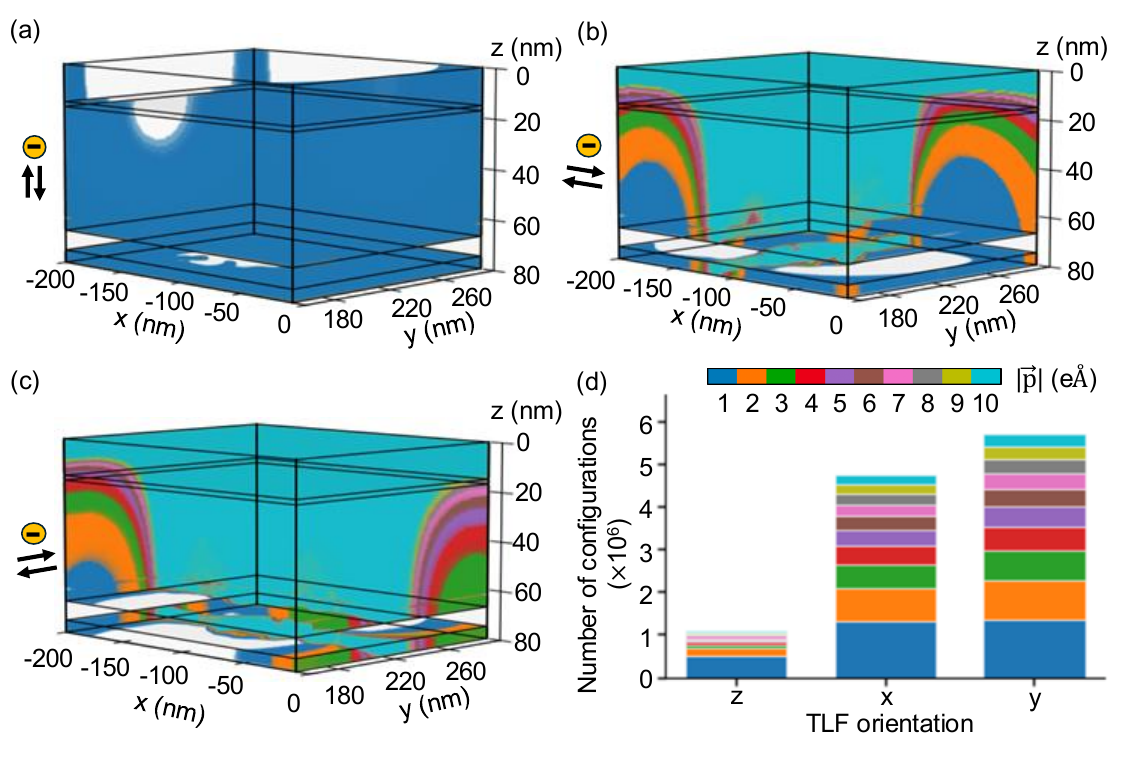}
    \caption[] { Plot of TLF configurations with gate-referred voltage shifts that are lower than the noise threshold and the distribution of TLF configurations across different dipole moments. (a)-(c) TLF configurations oriented along z, x and y directions respectively that register gate-referred voltage shifts of 20 $\mu$V and lower. The TLF configurations are color coded based on the magnitude of the dipole moment. The black lines indicate material boundaries. (d) The bar graph shows the distribution of TLF configurations plotted in (a)-(c) across different dipole moments for TLFs oriented along the z, x and y directions respectively.}
    \label{fig:noise_threshold}
\end{figure}

To quantitatively assess the agreement between the simulated and experimental results, we compared the simulated values of both the TLF-induced chemical-potential shift at the sensing dot and the associated detuning lever arm against the corresponding experimental measurements. A candidate fluctuator configuration was deemed consistent with experiment only if, for every measured observable, the simulation interval defined by the estimated numerical uncertainty overlapped the corresponding experimental confidence interval. The sources of uncertainty are discussed in Appendix~\ref{app:uncertainty_sources_quantification}. The surviving configurations define the localization region consistent with the full experimental signature.

\section{Results}

We varied the location, orientation and magnitude of displacement of the monopole and determined the configurations consistent with each dataset within 95 \% confidence. Based on the overlap of the configurations consistent with the various experimental measurements, we determine that the TLF is located in the ALD aluminum oxide. The TLF was localized to a region bounded by x = -157 nm and x = -156 nm, y = 230 nm and y = 261 nm, and z = 6 nm and z = 10 nm, as shown in Fig. \ref{fig:results_fig_device_layout}(c). Its lateral location is marked by a red arrow on Fig. \ref{fig:results_fig_device_layout}(d). The TLF-induced chemical potential shifts alone were insufficient to constrain the TLF location, as shown in Fig.~\ref{fig:results_fig_device_layout}(a) and Fig.~\ref{fig:progressive_constraints} (Appendix \ref{app:progressive_constraints}). Incorporating the gate-TLF detuning lever arms was critical to localizing the TLF, as shown in Fig.~\ref{fig:results_fig_device_layout}(b) and Fig.~\ref{fig:progressive_constraints}. The TLF was determined to be vertically oriented, with an electric dipole moment consistent with charge displacements of both 3 $\mathring{A}$ and 4 $\mathring{A}$.

To assess the robustness of the localization methodology, we generated several synthetic fluctuators spanning different locations, orientations, and electric dipole moment magnitudes. These representative synthetic fluctuators are shown in Fig. \ref{fig:validation_study}(a). In Fig. \ref{fig:validation_study}, we demonstrate localization for one of the vertically oriented synthetic fluctuators with an electric dipole moment of 6 e$\mathring{\mathrm{A}}$.  For each synthetic fluctuator, the corresponding sensor chemical-potential shift and gate-TLF detuning lever arms were generated from the simulated data and subsequently used as inputs to the localization procedure. In Fig. \ref{fig:validation_study}(b), uncertainties of 50 $\mu$V for the gate-referred voltage shift and 1 $\mu$eV/mV for the detuning lever arms were assumed. Within these uncertainty bounds, the synthetic fluctuator is confined to a relatively large volume. The high uncertainty data are consistent with dipole moments ranging from 3 to 10 e$\mathring{\mathrm{A}}$. The localization volume of the TLF configurations spans from z = 5 nm to 79 nm. For the vertically oriented fluctuators, the lateral extent of the admissible region in the $xy$-plane ranges from x = -179 nm to x = -65 nm and from y= 178 nm to y = 265 nm. Additionally, 0.04\% of the TLF configurations consistent with the high uncertainty data are horizontally oriented TLFs. Fig. \ref{fig:validation_study}(c) shows the localization obtained when the assumed uncertainties are reduced to 20 $\mu$V for the gate-referred voltage shift and 0.1 $\mu$eV/mV for the detuning lever arms. The tighter uncertainty bounds substantially improve the localization. Only vertically oriented TLF configurations remain consistent with the simulated measurements, all with a dipole moment magnitude of 6 e$\mathring{\mathrm{A}}$ except for one with a magnitude of 5 e$\mathring{\mathrm{A}}$. Furthermore, the localization volume contracts significantly, with the vertical range reducing to z = 16 nm to z = 38 nm and and the lateral extent in the $xy$ plane reducing to x = -160 nm to x = -158 nm and y = 195 nm to y = 200 nm. The localization results for the remaining fluctuators under the tighter uncertainty bounds are in Appendix \ref{app:remaining_synthetic_fluctuators}. This method is generally more effective at more precisely localizing vertically oriented fluctuators. We determined all TLF configurations that register a gate-referred voltage shift under the noise threshold of 20 $\mu$V. The locations of these fluctuators are plotted for three dipole moment orientations in Fig. \ref{fig:noise_threshold}. As shown, most of the vertically oriented fluctuators that measure a gate-referred voltage shift under the noise threshold are those with an electric dipole moment of 1 e$\mathring{A}$. For the horizontally oriented fluctuators, all electric dipole moments tested from 1 to 10 e$\mathring{A}$ register a signal under this threshold. This indicates that it is much more likely that a vertically oriented fluctuator is measured because they result in a stronger shift to the chemical potential of the sensing dot compared to horizontally oriented fluctuators of the same electric dipole moment. Additionally, among the synthetic TLFs tested, vertically oriented fluctuators exhibit, on average, larger gate-TLF detuning lever arms than their horizontally oriented counterparts. Consequently, fewer TLF configurations are consistent with the simulated data, enabling more precise localization of vertically oriented fluctuators.

\section{Conclusion}
We have developed a methodology for spatially localizing individual charge fluctuators in semiconductor quantum dot devices using measurements readily accessible in existing architectures. By combining fluctuator-induced chemical-potential shifts with gate–TLF detuning lever arms and self-consistent Schrodinger-Poisson modeling, we constrain the location of an experimentally observed fluctuator to within 1 nm, 31 nm, and 4 nm along the x, y, and z directions, respectively. The inclusion of gate–TLF detuning lever arms provides additional spatial constraints beyond those obtained from chemical-potential shifts alone. We further demonstrate the applicability of this approach across a range of fluctuator positions, electric dipole moments, and orientations, finding that localization is more effective for vertically oriented fluctuators due to their greater influence on the sensing dot's chemical potential and larger gate-TLF detuning lever arms. Additional measurements in devices with higher gate densities, or experiments involving deliberately positioned defects, could provide further validation of the inferred locations. More broadly, our results establish a framework for using semiconductor quantum dots as local probes of their electrostatic environment, offering a pathway toward identifying microscopic sources of charge noise and understanding their impact on qubit performance.

\begin{acknowledgments}
This material is based upon work supported by the Air Force Office of Scientific Research under award number FA9550-23-1-0710. Research was sponsored by the Army Research Office under Award number: W911NF-25-1-0141.The views and conclusions contained in this document are those of the authors and should not be interpreted as representing the official policies, either expressed or implied, of the Army Research Office or the U.S. Government.  The U.S. Government is authorized to reproduce and distribute reprints for Government purposes notwithstanding any copyright notation herein.  Additionally, we would like to thank Feiyang Ye, Tobias Jacobson and Dylan Albrecht for useful discussions. 
\end{acknowledgments}

\appendix


\nocite{*}

\bibliographystyle{apsrev4-2}
\bibliography{SpatialTriangulation}

\appendix
\setcounter{secnumdepth}{2}

\section{Ruling out charge traps}
\label{app:chargeTrap}

\begin{figure}[htbp]
    \centering
    \includegraphics[width=0.43\textwidth]{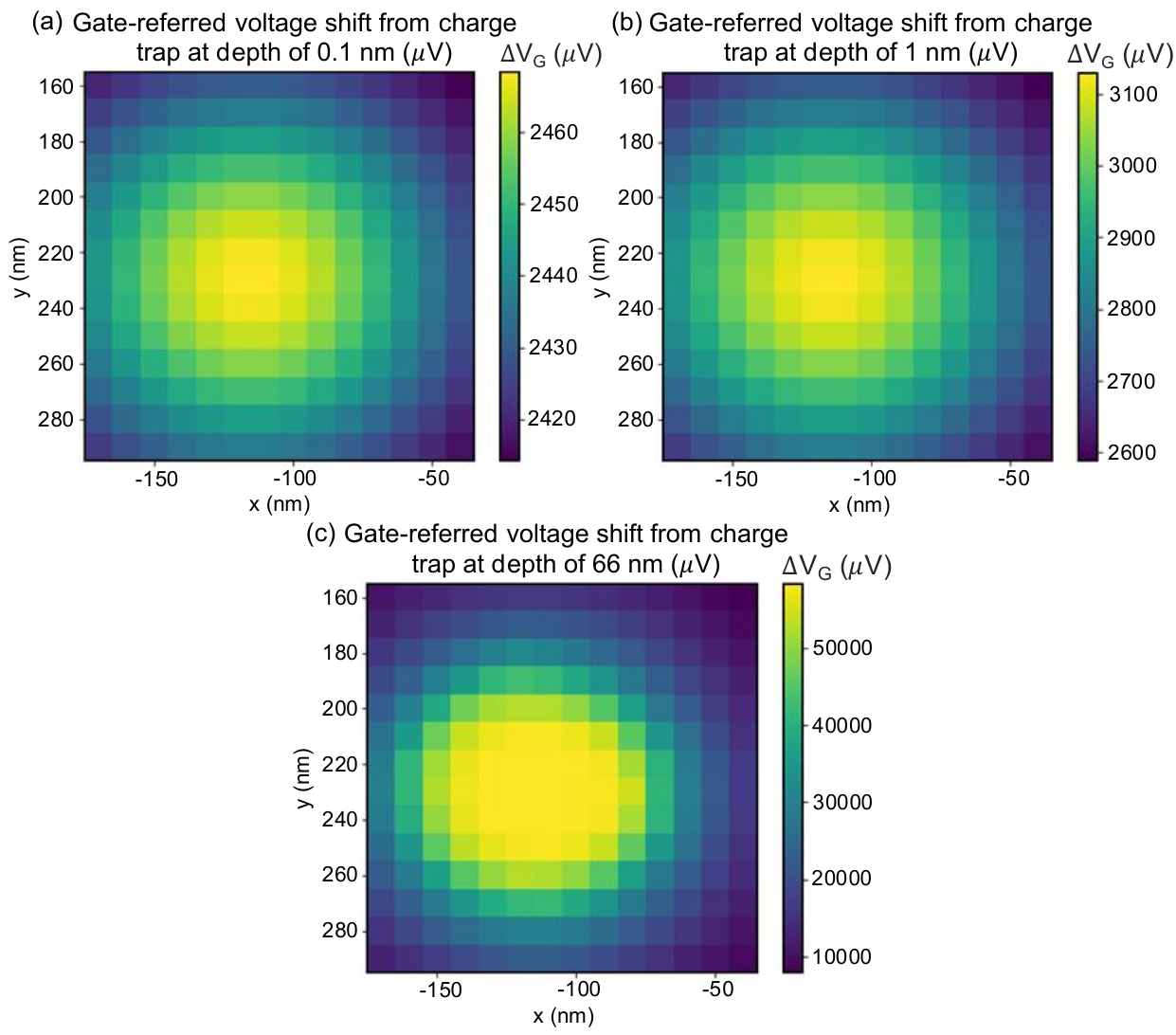}
    \caption[] { Gate-referred voltage shifts for charge traps placed at different heights. (a) Gate-referred voltage shift from charge trap placed 0.1 nm below the gate/oxide interface. (b) Gate-referred voltage shift from charge trap at 1 nm below gate/oxide interface. (c) Gate-referred voltage shift from charge trap located at a depth of 66 nm or 1 nm above the silicon well}
    \label{fig:charge_traps}
\end{figure}

In addition to bistable fluctuators, charge traps are another candidate model for defects. For a charge trap to mimic the behavior of a TLF, it must be within tunneling distance of a charge reservoir or metallic gate for electron exchange. Here, we investigated candidate trap locations 0.1 and 1 nm beneath the gates in the oxide as well as 1 nm above the quantum well. We determined that the simulations rule out charge traps at any of these locations because the gate-referred voltage shift of the charge trap is found to be at least 2.4 mV, which is an order of magnitude higher than the experimentally measured value of 200~$\mu$V\cite{ye_characterization_2024}, as shown in Fig. \ref{fig:charge_traps}.

\section{Sources of uncertainty}
\label{app:uncertainty_sources_quantification} 
\begin{figure}[htbp]
    \centering
    \includegraphics[width=0.47\textwidth]{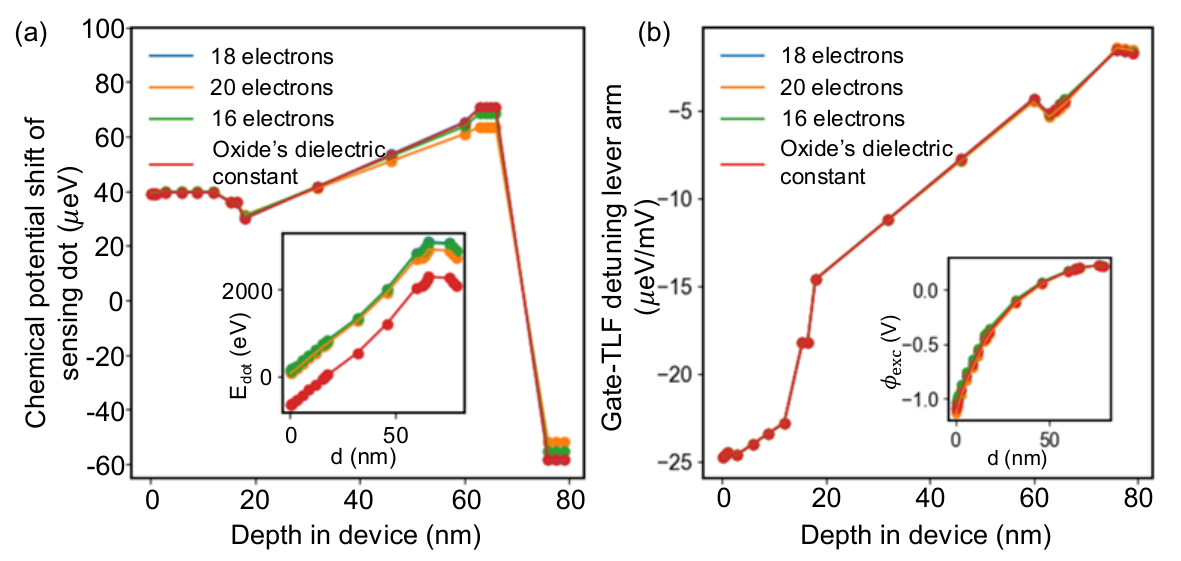}
    \caption[] {Sensitivity of the simulated observables to variations in electrostatic-model parameters as a function of TLF depth \(d\). (a) TLF-induced chemical-potential shift of the sensing dot calculated using 16, 18, and 20 electrons in the sensing dot and a different dielectric constant for the oxide. The inset shows the corresponding sensor dot chemical potential, \(E_{\mathrm{dot}}\).(b) Gate-TLF detuning lever arm for P obtained for the same model variations. The inset shows the excluded electrostatic potential, \(\phi_{\mathrm{exc}}\), used to calculate the gate-TLF detuning lever arm. }
    \label{fig:uncertainty_sources}
\end{figure}

Several sources of uncertainty enter the final triangulation results. The confidence intervals on the detuning lever arms extracted from experiment are reported in Ref.~\cite{ye_characterization_2024}. On the modeling side, two inputs are not precisely known: the dielectric constant of the ALD aluminum oxide and the number of electrons in the sensing dot. To quantify sensitivity to the oxide permittivity, we repeat the simulations with $\varepsilon_r = 8$ in place of the nominal $\varepsilon_r = 10$ used in the main calculations, and compare the TLF-induced chemical potential shift of the sensing dot and the gate--TLF detuning lever arm with respect to gate P along a linecut of TLF positions through the center of the sensing dot (Figs.~\ref{fig:uncertainty_sources}(a) and (b), respectively). Likewise, the sensing dot is estimated to contain 18 electrons; to account for uncertainty in this estimate, we repeat the calculations with 16 and 20 electrons [also shown in Figs.~\ref{fig:uncertainty_sources}(a) and (b)]. For each parameter, we take the maximum relative deviation from the nominal result along the linecut as the corresponding uncertainty and propagate it into the simulated detuning lever arms and gate-referred voltage shifts. Finally, we estimate the discretization error from a mesh-refinement test, which changes the sensing dot's chemical potential and excluded potential by 3\% and 0.1\% in the plane 1 nm above the quantum well; this error is included in the same way.

\section{Simulation methodology}
\label{app:sim_methodology}            

On the simulation side, accurate localization required modeling the electrostatic response to an angstrom-scale charge displacement in a realistic screened device environment. We model the TLF as a point charge switching between two closely spaced locations corresponding to the minima of the fluctuator potential. To compute the gate-TLF detuning lever arm, we use an excluded potential, defined as the potential at the location of the noise source with the self-potential of the fluctuating charge removed. Evaluating this quantity at both TLF positions allows us to account for changes in environmental screening as the charge switches state, an effect that becomes important near gates and charge reservoirs in the 2DEG. Several numerical developments were required to make this search both accurate and computationally tractable. We implemented local mesh refinement around the inserted charge to accurately resolve its electrostatic response. Precise localization of the TLF required evaluating the excluded potential and the sensing dot's chemical potential shift for TLFs placed on a finely spaced three-dimensional grid. Directly simulating every candidate location would have been prohibitively expensive. Since the simulated observables varied smoothly as a function of TLF position, we used higher-order spline interpolation to efficiently sample the large parameter space while retaining the accuracy needed to resolve angstrom-scale charge motion. These simulation advances made it possible to compare experimental measurements against a large parameter space of TLF positions, orientations, and dipole magnitudes with sufficient accuracy to identify a unique surviving spatial cluster as shown in Fig. \ref{fig:results_fig_device_layout}(c).

\section{Synthetic fluctuators}
\label{app:remaining_synthetic_fluctuators}

\begin{figure}[htbp]
    \centering
    \includegraphics[width=0.5\textwidth]{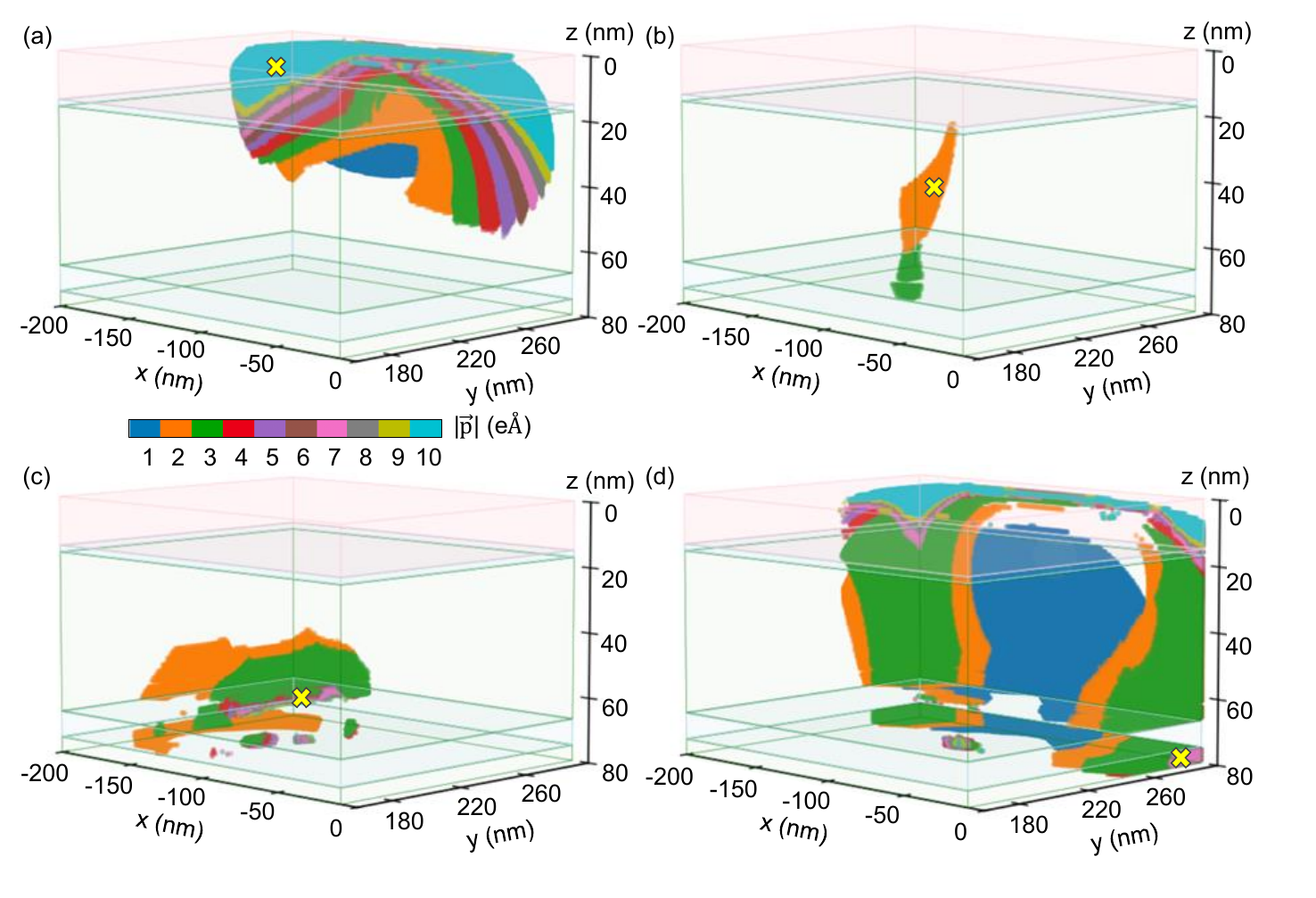}
    \caption{Localization of simulated fluctuators using the spatial triangulation methodology. Each panel shows the TLF configurations consistent with all simulated experimental signatures, with color indicating the magnitude of the electric dipole moment. The true location of each simulated TLF is marked by a yellow cross. (a) Configurations consistent with the \(y\)-oriented TLF with an electric dipole moment of \(7~e\text{\AA}\), indicated by the pink arrow in Fig.~\ref{fig:uncertainty_sources}(a). (b) Configurations consistent with the vertically oriented TLF with an electric dipole moment of \(2~e\text{\AA}\), indicated by the orange arrow. (c) Configurations consistent with the \(x\)-oriented TLF with an electric dipole moment of \(9~e\text{\AA}\), indicated by the yellow arrow. (d) Configurations consistent with the \(y\)-oriented TLF with an electric dipole moment of \(4~e\text{\AA}\), indicated by the red arrow.}
    \label{fig:synthetic_fluctuators}
\end{figure}

The localization for the fluctuators depicted in Fig. \ref{fig:validation_study}(a) are shown in Fig. \ref{fig:synthetic_fluctuators}. The localization volumes plotted in Fig. \ref{fig:synthetic_fluctuators}(a)-(d) correspond to fluctuators with dipole moments 7, 2, 9 and 4 e$\text{\AA}$ respectively. These 4 fluctuators are oriented along the y, z, x and y directions respectively. The localization volumes are large for the fluctuators shown in Fig. \ref{fig:synthetic_fluctuators}(a) and (d). While most of the orientations of the shortlisted configurations for Fluctuators 1, 3 and 4 agree with the actual TLF orientation, there are a handful of the  TLF configurations that have a different orientation. Additionally, all electric dipole moments tested are represented in the shortlisted TLF configurations for all horizontally oriented fluctuators tested as shown in Fig. \ref{fig:synthetic_fluctuators}(a),(c) and (d). The localization generally works better for vertically oriented fluctuators as shown in Fig. \ref{fig:synthetic_fluctuators}(b) and Fig. \ref{fig:uncertainty_sources}(b) due to their larger gate-TLF detuning lever arms compared to the horizontally oriented fluctuators.

\section{Progressive Constraints on TLF Configurations}
\label{app:progressive_constraints}

\begin{figure}[htbp]
    \centering
    \includegraphics[width=0.47\textwidth]{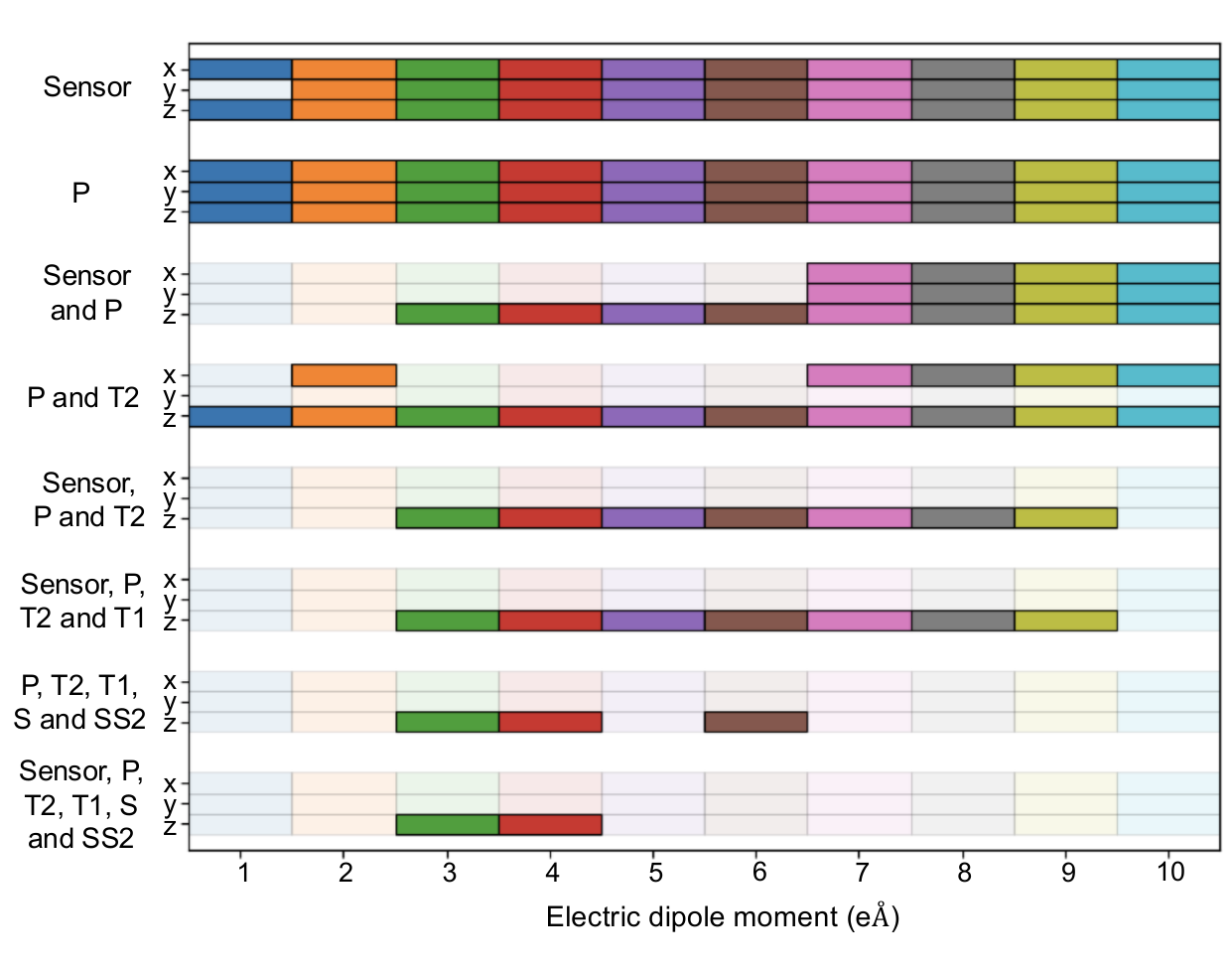}
    \caption{ Experimentally consistent TLF configurations as successive constraints are applied. Colored cells indicate electric dipole moments and orientations (x, y, and z) of TLF configurations consistent with the indicated experimental observables, while faint cells denote configurations that are excluded. The x and y orientations correspond to horizontally oriented TLFs, while the z orientation corresponds to vertically oriented TLFs. The gate-referred voltage shift (denoted by `Sensor') and P detuning lever arm data individually permit a broad range of TLF configurations, while combining multiple observables progressively constrains the allowed dipole moments and orientations. Applying all measured constraints leaves only a vertically-oriented TLF with an electric dipole moment of 3 and 4 e\AA.
}
    \label{fig:progressive_constraints}
\end{figure}

Fig. \ref{fig:progressive_constraints} shows the results of including different datasets in the final localization. The gate-referred voltage shift dataset is referred to as 'Sensor' and the gates with respect to which detuning lever arms were determined are indicated by the respective gate name.

\end{document}